\documentclass[a4paper,11pt]{article}
\usepackage{pos}
\usepackage{hyperref}

\title{Hybrid concept of detection for a wide-field gamma-ray observatory using Cherenkov telescopes}
\ShortTitle{Cherenkov telescopes for wide-field gamma-ray observatory}

\author*[a]{Alena Bakalov\'a}
\author[b]{, Ruben Concei\c{c}\~{a}o}
\author[b]{, Lucio Gibilisco}
\author[b]{, Mario Pimenta}
\author[b]{, Bernardo Tom\'e}
\author[a]{ and Patrik \v{C}echvala}
\author[a,c]{, Vladim\'{i}r Novotn\'{y}}
\author[a]{, Jakub V\'{i}cha}
\author[a]{, Jakub Jury\v{s}ek}

\manuallySeparateAuthors
\forColl{SST-1M}

\affiliation[a]{Institute of Physics of the Czech Academy of Sciences, Prague, Czech Republic}

\affiliation[b]{Laboratory of Instrumentation and Experimental Particle Physics, Lisbon, Portugal}

\affiliation[c]{Institute of Particle and Nuclear Physics, Faculty of Mathematics and Physics, Charles~University,\\
  V Hole\v sovi\v ck\'ach 2, 180 00 Prague, Czech Republic}

\emailAdd{bakalova@fzu.cz}

\abstract{The hybrid detection approach in astroparticle physics has been successfully employed in cosmic-ray experiments and is currently being explored by gamma-ray observatories like LHAASO. We present a study on the hybrid detection concept for the future Southern Wide-field Gamma-ray Observatory (SWGO), integrating multiple Cherenkov telescopes represented in the analysis by Single-Mirror Small-Size imaging atmospheric Cherenkov Telescopes (SST-1M) located next to the surface array of water Cherenkov detectors (WCDs). We discuss the mutual benefits of this hybrid approach and present simulation-based results on key performances. Our findings points to the fact that the combination of wide field-of-view and continuous operation of WCDs with the high angular and energy resolution of Cherenkov telescopes could significantly improve the overall detection capabilities of the SWGO experiment.}

\ConferenceLogo{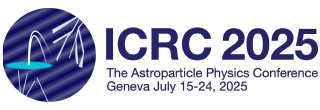}

\FullConference{39th International Cosmic Ray Conference (ICRC2025)\\
 15–24 July 2025\\
Geneva, Switzerland\\}

\begin{document}
\maketitle

\section{Introduction}

While existing wide-field observatories of gamma rays, such as HAWC and LHAASO, are providing valuable information about the gamma ray sources in the Northern sky, no such experiment currently operates in the Southern hemisphere. The Southern Wide-field Gamma-ray Observatory (SWGO) \cite{SWGO} is a proposed next generation experiment designed to fill in this observational gap. The information that can be obtained from the continuous wide-field observations is a powerful tool to study gamma rays, but it has been demonstrated that a hybrid detection approach, combining measurements from Imaging Air Cherenkov Telescopes (IACT) with a surface detector array, can enhance the overall detection performance (see e.g.~\cite{LHAASOhybrid}).  

As a first step towards the full hybrid detection approach, we study how a wide-field array of surface detectors can contribute to improving the operation and data analysis of SST-1M Cherenkov telescopes \cite{SST-1M} placed inside the surface array. We perform a simulation-based analysis of a configuration with two identical\footnote{Hardware configuration of the more recent prototype SST-1M-2 was used \cite{Alispach:2025mbz}.} SST-1M telescopes working in stereo mode and present preliminary results on the key performances, including the gamma/hadron ($\gamma/h$) separation, flux sensitivity, and energy and angular resolution. The analysis follows a reconstruction pipeline similar to the standard SST-1M reconstruction procedure \cite{Alispach:2025mbz} with additional $\gamma/h$ discrimination parameters $LCm$ and $P^{\alpha}_{\rm{tail}}$ obtained from the array of surface detectors. 

\section{Method}

\subsection{Simulation settings}

The presented study is based on three-step Monte Carlo (MC) simulations.
First, particle interactions and Cherenkov light from air showers were simulated in \texttt{CORSIKA v7.7402}~\cite{1998cmcc.book.....H} employing as hadronic interaction models for low- and high-energy interactions \texttt{UrQMD}~\cite{urqmd2} and \texttt{QGSJet\,II-04}~\cite{qgs}, respectively.
Second, the attenuation of the light and the response of the SST-1M telescopes was calculated in \texttt{sim\_telarray v2021-12-25}~\cite{BERNLOHR2008149}.
Finally, the response of the ground array to the particles produced in the first step was evaluated using the simplified parametric code~\cite{Conceicao:2022lkc_LCm}.

\begin{figure}
    \centering
    \includegraphics[width=0.7\linewidth]{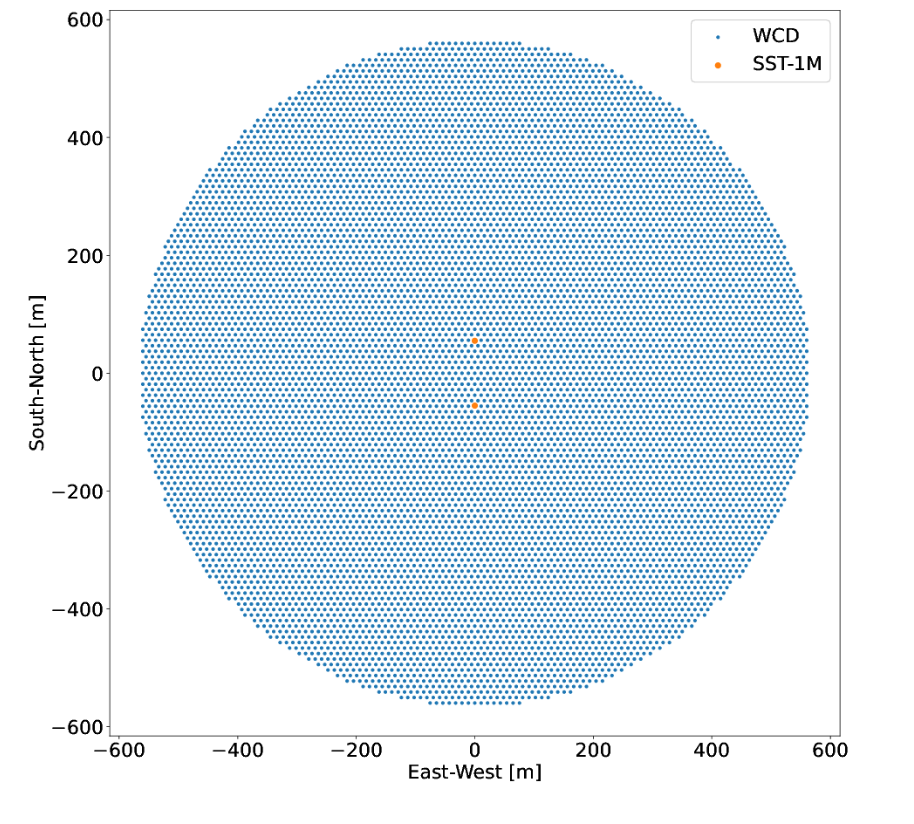}
    \caption{Sketch of the array of WCDs together with positions of 2 SST-1Ms.}
    \label{fig:array}
\end{figure}

The ground array, shown in Fig.~\ref{fig:array}, consisted of 9997 water-Cherenkov detectors (WCD), each covering 12.6\,m$^2$, spread over an area of 1\,km$^2$ and equipped with 3~PMTs.
The fill factor of the array was~12.5\%.
The simulated altitude was 4700\,m and the atmospheric density profile, together with the geomagnetic field, was adjusted to the Pampa la Bola site located in the Atacama Astronomical Park, Chile.
Although the geographical position is tailored to the future SWGO site~\cite{SWGO:2025taj}, main conclusions can be generalized to other high-altitude sites as well.

The SST-1M simulations were performed for low night-sky background with a photon rate of 72\,MHz, the corresponding response of pixels was tuned to those conditions \cite{Alispach:2025mbz}.
The atmosphere transmissivity was calculated using MODTRAN~\cite{modtran} for Pampa la Bola.
Pointing direction of the telescopes was 20$^{\circ}$ zenith angle and showers came from the North to the South.
The positions of the two SST-1Ms which were 110\,m apart are depicted in orange in Fig.~\ref{fig:array}.

Monte Carlo sets of diffuse gamma rays, protons, and point-like gamma rays were simulated to train the Random Forests (RF) used in the reconstruction of showers in \texttt{sst1mpipe}~\cite{jurysek_2025_14808846}, which is an open source software developed for calibration and shower reconstruction for SST-1M, and to directly evaluate the sensitivity to point-like source of gamma rays, respectively.
The energy of the primaries was between 200\,GeV and 631\,TeV, and 400\,GeV and 1100\,TeV for gamma rays and protons, respectively, both simulated according to $\frac{\rm{d}N}{\rm{d}E} \propto E^{-2}$.
In the case of diffuse samples, the opening angle of $10^{\circ}$ was used to cover the entire field of view of SST-1M.
Showers were thrown up to the impact distance of 1032\,m from the ground array center. However, during the analysis, events with shower core positions above 565\,m, the radius of the array, were discarded from the ground array-enhanced analysis, in order to explicitly confirm that the array size is sufficient to well cover the effective area of the SST-1M telescopes (only $\sim2\%$ of reconstructed gammas were discarded).

\subsection{$LCm$ and $P^{\alpha}_{\rm{tail}}$}

As mentioned above, the signal from the surface detector array was evaluated using a simplified code from \cite{Conceicao:2022lkc_LCm}. The signal in the WCDs is calculated from the shower footprint at the ground level (4700 m a.s.l.) obtained from CORSIKA simulations. Two $\gamma/h$ discrimination parameters, which can be easily obtained from an array of WCDs, are extracted from the simulated response of the ground array. For comparison, we also use the true number of muons $N_{\mu}$ from the CORSIKA simulations as an ideal reference case.

The first parameter obtained from the ground array is $LCm$, a $\gamma/h$ discriminator first introduced in \cite{Conceicao:2022lkc_LCm}. The $LCm$ variable characterizes the azimuthal asymmetry of the ground-level shower footprint and has been shown to be a powerful discriminator between gamma and hadron-induced showers \cite{Conceicao:2022lkc_LCm, Conceicao:2023LCm, LCmPeV}. The second $\gamma/h$ discriminator obtained from the surface array is the $P^{\alpha}_{\rm{tail}}$ \cite{Conceicao:2023ybu_Ptail}. This variable is built from the total signal detected by the ground array of WCDs and shows a strong correlation with the total number of muons in the shower.

Both parameters, $LCm$ and $P^{\alpha}_{\rm{tail}}$, are evaluated only for showers with energy above 10\,TeV, which is the lower energy limit where these parameters can be evaluated with the software used and below which the meaning of the parameters starts to be problematic.
The true number of muons is extracted for showers at all energies. The surface-array parameters were added to the features used for training of RF classifier for $\gamma/h$ separation in \texttt{sst1mpipe}.

\section{Results}

\begin{figure}
    \centering
    \includegraphics[width=0.6\linewidth]{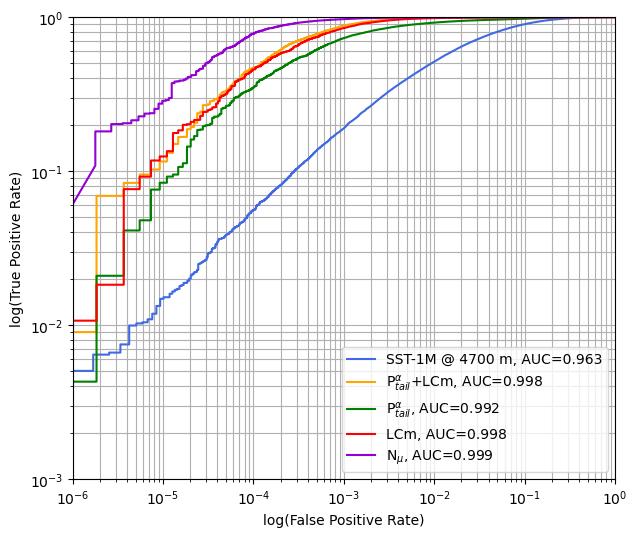}
    \caption{Receiver Operating Characteristics. Solely for SST-1M telescopes (blue) and for SST-1M telescopes with $\gamma/h$ discrimination parameters obtained from the ground array added into the \texttt{sst1mpipe}, including the $P^{\alpha}_{\rm{tail}}$ (green), $LCm$ (red), both $P^{\alpha}_{\rm{tail}}$ and $LCm$ (orange) and true number of muons $N_{\mu}$ (purple).  Those of $LCm$ and $P^{\alpha}_{\rm{tail}}$ are shown for energies above 10\,TeV only.}
    \label{fig:roc}
\end{figure}

The most significant improvement is obtained for the $\gamma/h$ separation performance. This is demonstrated in Figure~\ref{fig:roc}, which shows the receiver operating characteristics (ROC). The performance of the $\gamma/h$ classifier is expressed as the integral of the ROC curve, area under the curve (AUC). With the addition of the $LCm$ and $P^{\alpha}_{\rm{tail}}$ variables from the ground array, the achieved AUC is 0.998 compared to 0.963 from the telescopes only.

\begin{figure}
    \centering
    \includegraphics[width=1\linewidth]{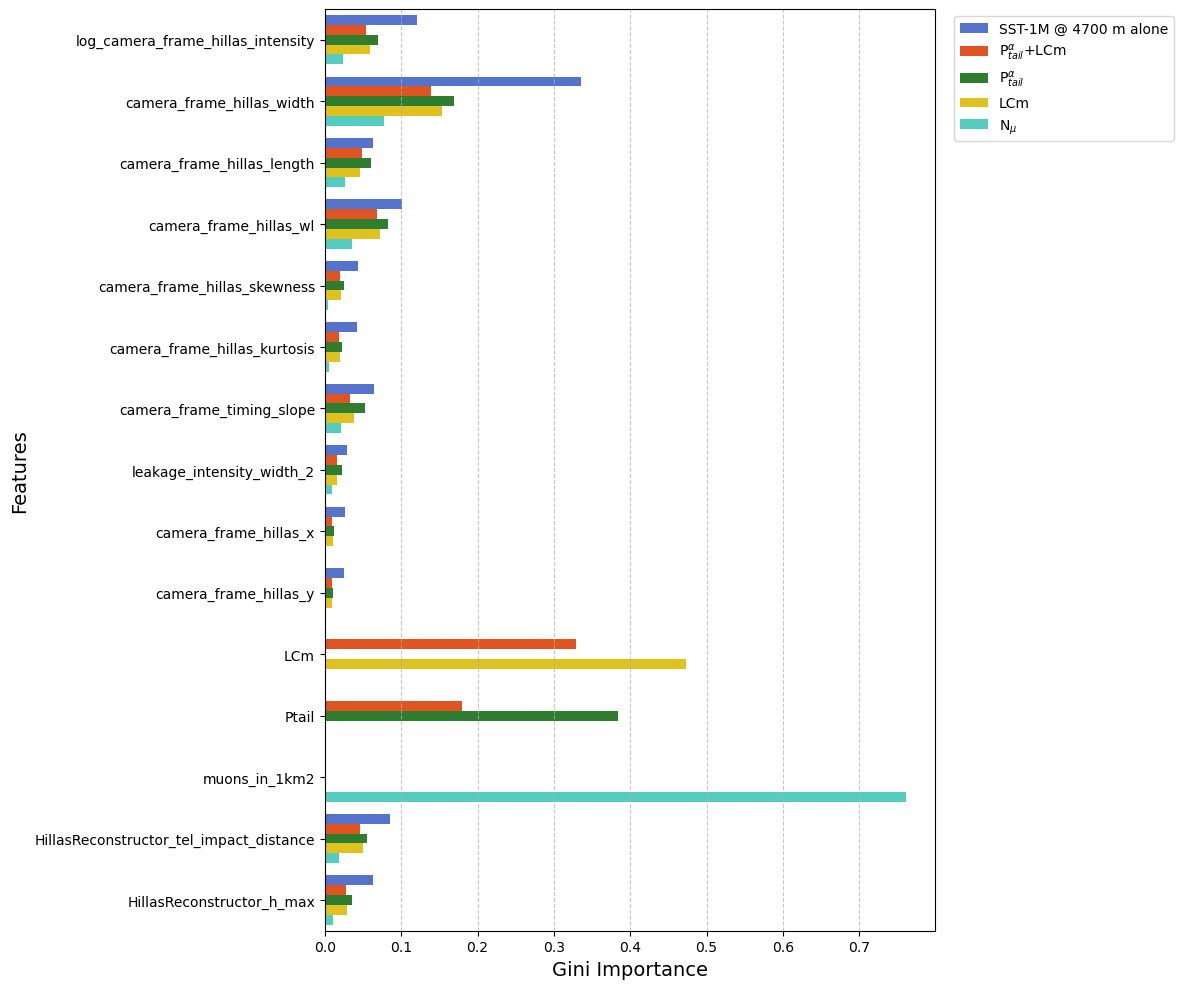}
    \caption{The Gini importance for $\gamma/h$ Random Forests classifier of the sole SST-1M telescopes (blue) and of SST-1M telescopes with the additional information from the ground array: true N$_{\mu}$ (muons\_in\_1km2, cyan), LCm (yellow), P$^{\alpha}_{\rm{tail}}$ (green) and P$^{\alpha}_{\rm{tail}}$+LCm (red). For the sole SST-1M telescopes the standard features which are used for the reconstruction: log of the intensity, width, length, width/length ratio, skewness, kurtosis, timing slope, leakage, coordinates of the shower center of gravity in the FoV (x,y). The parameters derived from stereo reconstruction the distance of the impact point from the telescope (impact dist) and the height of the shower maximum (h$_{max}$) were also used.  }
    \label{fig:importance}
\end{figure}

The Gini importance\footnote{The Gini importance is a measure of how much individual features contribute to the decision-making process of the Random Forrest. For more details, see \url{https://scikit-learn.org/stable/modules/tree.html}.} of individual reconstruction parameters for different $\gamma/h$ classifiers is shown in Figure~\ref{fig:importance}.
Different colors correspond to the sole SST-1M telescopes (blue) and analyses enhanced by ground-array information.
In the case of pure SST-1M reconstruction, the largest importance exhibits \texttt{camera\_frame\_hillas\_width}, a measure of the spread of a shower on camera, which is well known $\gamma/h$ discriminator of IACTs~\cite{Alispach:2025mbz}.
When the true number of muons is included, the decision is heavily based on it, demonstrating the superior separation power of $N_{\mu}$.
The interpretation of $P^{\alpha}_{\rm{tail}}$ and $LCm$ is slightly more complicated, because they contribute only at $E>10\,\mathrm{TeV}$, while the figure shows importance over the full energy range.
Nevertheless, both $P^{\alpha}_{\rm{tail}}$ and $LCm$ overall dominate the decision logic when used, and for their combination, $LCm$ contributes more than $P^{\alpha}_{\rm{tail}}$, showing its better $\gamma/h$ discrimination power at ultra-high energies~\cite{Conceicao:2023LCm,Conceicao:2023ybu_Ptail,Conceicao:2022lkc_LCm}.

\begin{figure}
    \centering
    \includegraphics[width=0.6\linewidth]{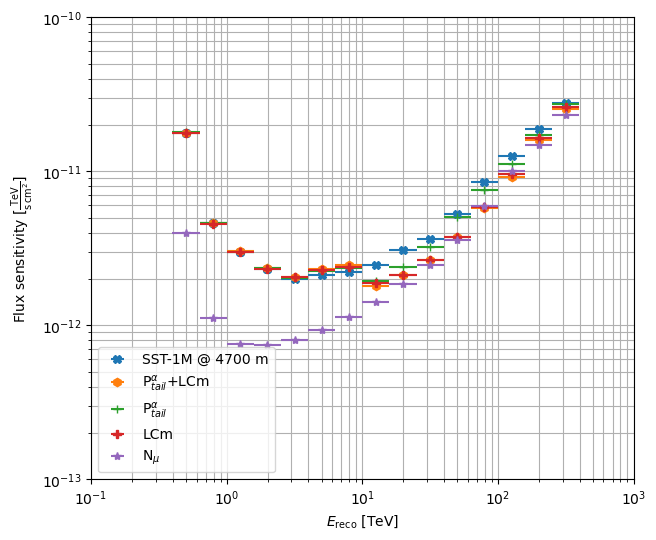}
    \caption{The flux sensitivity for the sole SST-1M telescopes (blue) and SST-1M telescopes with the additional information from the ground array (same as in Figure~\ref{fig:roc}).}
    \label{fig:sensitivity}
\end{figure}

Figure~\ref{fig:sensitivity} shows the effect of the above-mentioned performance on the flux sensitivity of the telescopes. Incorporating the parameters from the ground array into the reconstruction improves the flux sensitivity by $\sim30\%$ above 10 TeV. The ideal case with the use of $N_{\mu}$ would increase the sensitivity over the whole energy range.
The fact that $LCm$ even surpasses in efficiency $N_\mu$ \cite{Conceicao:2023LCm} is visible in the energy bins around 100\,TeV for which the calculation of $LCm$ was optimized.
In other bins $N_\mu$ works better, which is unexpected, although the RF classification may result in slightly different outputs than those investigated in \cite{Conceicao:2023LCm}.
However, these differences will be the subject of further investigation.

\begin{figure}
    \centering
    \includegraphics[width=0.6\linewidth]{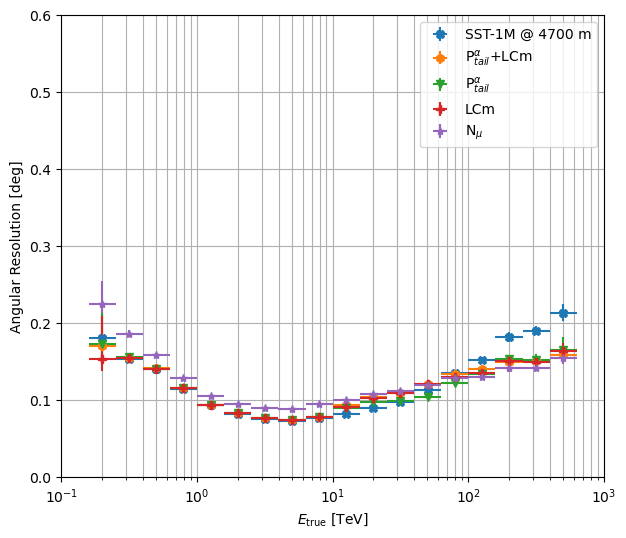}
    \caption{Angular resolution defined as 68\% containment as function of true energy for point-like source gammas of the SST-1M telescopes (blue) and SST-1M telescopes with the additional $\gamma/h$ separation parameters from the ground array (same as in Figure~\ref{fig:roc}). }
    \label{fig:ang_res}
\end{figure}

Interestingly, including the $\gamma$/h separation variables into the RF indirectly influences also the angular resolution, energy bias and energy resolution, but this is caused solely due to the selection effect that increases or decreases the amount of showers which pass the gamaness cut. Figure~\ref{fig:ang_res} shows the angular resolution of the SST-1M telescopes without and with the additional parameters from the ground array. In this case, the selection effect of including $N_{\mu}$ parameter causes worsening of the angular resolution at energies below few tens of TeV due to the incorporation of gamma events with poorer directional reconstruction.

\begin{figure}
    \centering
    \includegraphics[width=0.49\linewidth]{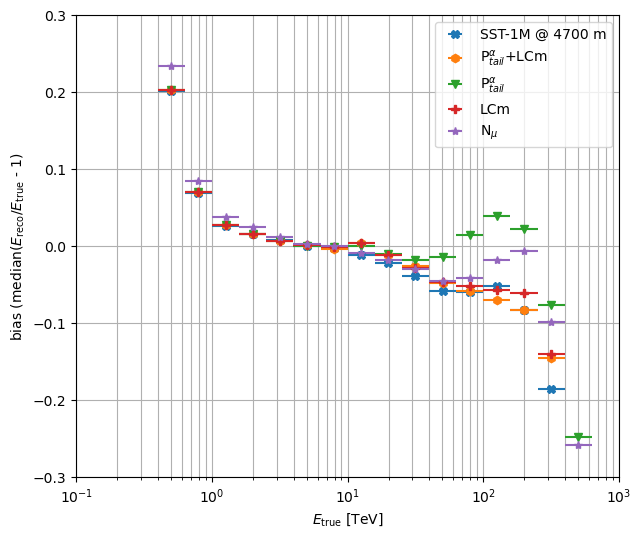}
    \includegraphics[width=0.49\linewidth]{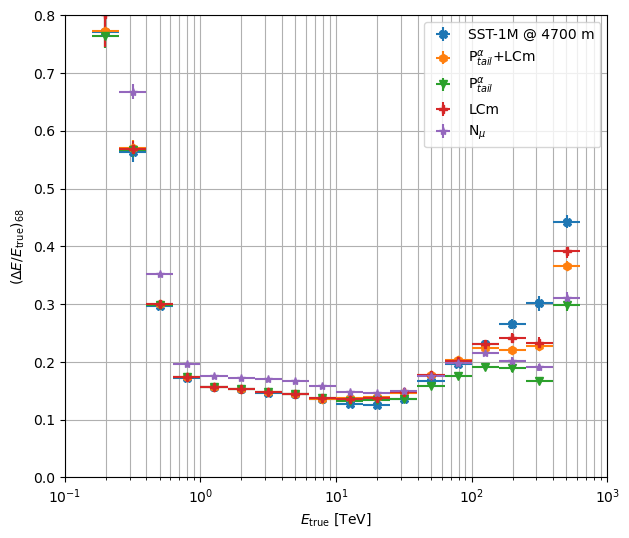}
    \caption{The energy bias (left) and energy resolution (right) for the sole SST-1M telescopes (blue) and SST-1M telescopes with the additional $\gamma/h$ separation parameters from the ground array (same as in Figure~\ref{fig:roc}).}
    \label{fig:energy_bias_res}
\end{figure}

The energy bias and energy resolution are shown in Figure~\ref{fig:energy_bias_res}. The energy resolution degrades at energies above $\sim300$\,TeV, which is partly related to the altitude of the telescope position site. This performance might be improved with additional information from the ground array, including the reconstructed shower core position and energy estimator.




\section{Summary}

In this study, we investigate the performance of a hybrid detection of gamma-rays combining two SST-1M imaging atmospheric Cherenkov telescopes with a dense surface array of water-Cherenkov detectors, similar to that proposed for SWGO. Using a simulation-based approach, we evaluate the performance of the angular and energy resolution, $\gamma/h$ separation, and flux sensitivity of the SST-1M telescopes when taking into account $\gamma/h$ separation variables obtained from the ground array, $LCm$ and $P^{\alpha}_{\rm{tail}}$. We demonstrate that the inclusion of such parameters improves the reconstruction performance. Especially the $\gamma/h$ separation power is significantly increased and, consequently, the flux sensitivity of the telescopes is improved by $\sim30\%$ above 10\,TeV. 

We note, that these results are preliminary and only the $\gamma/h$ discriminators from the ground array are taken into account. In future work, using more thorough analysis and simulation of the whole surface detector array response and reconstruction, additional improvements of the telescope performance might be achieved by incorporating more information from the ground array, such as the reconstructed shower core position and energy. Moreover, the information from the telescopes, with their precise angular and energy resolution might help with the energy calibration of the surface array.

\acknowledgments


This publication was created as part of the projects funded in Poland by the Minister of Science based on agreements number 2024/WK/03 and DIR/\-WK/2017/12. The construction, calibration, software control and support for operation of the SST-1M cameras is supported by SNF (grants CRSII2\_141877, 20FL21\_154221, CRSII2\_160830, \_166913, 200021-231799), by the Boninchi Foundation and by the Université de Genève, Faculté de Sciences, Département de Physique Nucléaire et Corpusculaire. The Czech partner institutions acknowledge support of the infrastructure and research projects by Ministry of Education, Youth and Sports of the Czech Republic (MEYS) and the European Union funds (EU), MEYS LM2023047, EU/MEYS CZ.02.01.01/00/22\_008/0004632, CZ.02.01.01/00/22\_010/0008598, Co-funded by the European Union (Physics for Future – Grant Agreement No. 101081515), and Czech Science Foundation, GACR 23-05827S.
The Portuguese contribution was supported by FCT - Funda\c{c}\~{a}o para a Ci\^encia e a Tecnologia, I.P. by PRT/BD/154192/2022 [\href{https://doi.org/10.54499/PRT/BD/154192/2022}{DOI}].

{\footnotesize
\raggedright
\setlength{\parskip}{0ex}
\setlength{\itemsep}{0pt}
\bibliography{bibfile}{}
\bibliographystyle{JHEP_mod}
}



\clearpage
\section*{Full Authors List: SST-1M Collaboration}
\scriptsize
\noindent
C.~Alispach$^1$,
A.~Araudo$^2$,
M.~Balbo$^1$,
V.~Beshley$^3$,
J.~Bla\v{z}ek$^2$,
J.~Borkowski$^4$,
S.~Boula$^5$,
T.~Bulik$^6$,
F.~Cadoux$^`$,
S.~Casanova$^5$,
A.~Christov$^2$,
J.~Chudoba$^2$,
L.~Chytka$^7$,
P.~\v{C}echvala$^2$,
P.~D\v{e}dic$^2$,
D.~della Volpe$^1$,
Y.~Favre$^1$,
M.~Garczarczyk$^8$,
L.~Gibaud$^9$,
T.~Gieras$^5$,
E.~G{\l}owacki$^9$,
P.~Hamal$^7$,
M.~Heller$^1$,
M.~Hrabovsk\'y$^7$,
P.~Jane\v{c}ek$^2$,
M.~Jel\'inek$^{10}$,
V.~J\'ilek$^7$,
J.~Jury\v{s}ek$^2$,
V.~Karas$^{11}$,
B.~Lacave$^1$,
E.~Lyard$^{12}$,
E.~Mach$^5$,
D.~Mand\'at$^2$,
W.~Marek$^5$,
S.~Michal$^7$,
J.~Micha{\l}owski$^5$,
M.~Miro\'n$^9$,
R.~Moderski$^4$,
T.~Montaruli$^1$,
A.~Muraczewski$^4$,
S.~R.~Muthyala$^2$,
A.~L.~Müller$^2$,
A.~Nagai$^1$,
K.~Nalewajski$^5$,
D.~Neise$^{13}$,
J.~Niemiec$^5$,
M.~Niko{\l}ajuk$^9$,
V.~Novotn\'y$^{2,14}$,
M.~Ostrowski$^{15}$,
M.~Palatka$^2$,
M.~Pech$^2$,
M.~Prouza$^2$,
P.~Schovanek$^2$,
V.~Sliusar$^{12}$,
{\L}.~Stawarz$^{15}$,
R.~Sternberger$^8$,
M.~Stodulska$^1$,
J.~\'{S}wierblewski$^5$,
P.~\'{S}wierk$^5$,
J.~\v{S}trobl$^{10}$,
T.~Tavernier$^2$,
P.~Tr\'avn\'i\v{c}ek$^2$,
I.~Troyano Pujadas$^1$,
J.~V\'icha$^2$,
R.~Walter$^{12}$,
K.~Zi{\c e}tara$^{15}$ \\

\noindent
$^1$D\'epartement de Physique Nucl\'eaire, Facult\'e de Sciences, Universit\'e de Gen\`eve, 24 Quai Ernest Ansermet, CH-1205 Gen\`eve, Switzerland.
$^2$FZU - Institute of Physics of the Czech Academy of Sciences, Na Slovance 1999/2, Prague 8, Czech Republic.
$^3$Pidstryhach Institute for Applied Problems of Mechanics and Mathematics, National Academy of Sciences of Ukraine, 3-b Naukova St., 79060, Lviv, Ukraine.
$^4$Nicolaus Copernicus Astronomical Center, Polish Academy of Sciences, ul. Bartycka 18, 00-716 Warsaw, Poland.
$^5$Institute of Nuclear Physics, Polish Academy of Sciences, PL-31342 Krakow, Poland.
$^6$Astronomical Observatory, University of Warsaw, Al. Ujazdowskie 4, 00-478 Warsaw, Poland.
$^7$Palack\'y University Olomouc, Faculty of Science, 17. listopadu 50, Olomouc, Czech Republic.
$^8$Deutsches Elektronen-Synchrotron (DESY) Platanenallee 6, D-15738 Zeuthen, Germany.
$^9$Faculty of Physics, University of Bia{\l}ystok, ul. K. Cio{\l}kowskiego 1L, 15-245 Bia{\l}ystok, Poland.
$^{10}$Astronomical Institute of the Czech Academy of Sciences, Fri\v{c}ova~298, CZ-25165 Ond\v{r}ejov, Czech Republic.
$^{11}$Astronomical Institute of the Czech Academy of Sciences, Bo\v{c}n\'i~II 1401, CZ-14100 Prague, Czech Republic.
$^{12}$D\'epartement d'Astronomie, Facult\'e de Science, Universit\'e de Gen\`eve, Chemin d'Ecogia 16, CH-1290 Versoix, Switzerland.
$^{13}$ETH Zurich, Institute for Particle Physics and Astrophysics, Otto-Stern-Weg 5, 8093 Zurich, Switzerland.
$^{14}$Institute of Particle and Nuclear Physics, Faculty of Mathematics and Physics, Charles University, V Hole\v sovi\v ck\' ach 2, Prague 8, Czech~Republic.
$^{15}$Astronomical Observatory, Jagiellonian University, ul. Orla 171, 30-244 Krakow, Poland.

\end{document}